\documentclass[aps,prd,twocolumn,showpacs,nofootinbib]{revtex4-1}

\usepackage{amsfonts,color,graphicx,epsfig,amsmath,multirow}

\usepackage[colorlinks=true,linkcolor=blue,citecolor=blue, urlcolor=blue]{hyperref}

\begin{document}

\title{Photoproduction of doubly heavy baryon at the ILC}

\author{Gu Chen}
\author{Xing-Gang Wu}
\email{email:wuxg@cqu.edu.cn}
\author{Zhan Sun}
\author{Yang Ma}
\author{Hai-Bing Fu}

\address{Department of Physics, Chongqing University, Chongqing 401331, P.R. China}

\date{\today}

\begin{abstract}

In the present paper, we make a detailed study on the doubly heavy baryon photoproduction in the future $e^+e^-$ International Linear Collider (ILC). The baryons $\Xi_{cc}$, $\Xi_{bc}$, and $\Xi_{bb}$ are produced via the channel $\gamma \gamma \to \Xi_{QQ'} +\bar{Q'} +\bar{Q}$, where $Q$ and $Q'$ stand for heavy $c$ or $b$ quark, respectively. As for the $\Xi_{QQ'}$-baryon production, it shall first generate a $(QQ')[n]$-diquark and then form the final baryon via fragmentation, where $[n]$ stands for the color- and spin- configurations for the $(QQ')$-diquark states. According to the non-relativistic QCD theory, four diquark configurations shall provide sizable contributions to the baryon production, e.g., $[n]$ equals $[^3S_1]_{\bar{\textbf{3}}}$, $[^1S_0]_{\textbf{6}}$, $[^3S_1]_{\textbf{6}}$, or $[^1S_0]_{\bar{\textbf{3}}}$, respectively. We adopt the improved helicity amplitude approach for the hard scattering amplitude to improve the calculation efficiency. Total and differential cross sections of those channels, as well as the theoretical uncertainties, are presented. We show that sizable amounts of baryon events can be generated at the ILC, i.e., about $2.0\times 10^{6}$ $\Xi_{cc}$, $2.2\times 10^{5}$ $\Xi_{bc}$, as well as $3.0\times 10^{3}\;\Xi_{bb}$ events are to be generated in one operation year for $\sqrt{S}=500$ GeV and ${\cal L}\simeq 10^{36}$cm$^{-2}$s$^{-1}$.

\end{abstract}


\maketitle

\section{Introduction}

Many theoretical predictions for the production of doubly heavy baryons $\Xi_{QQ^{\prime}}$ have been done in Refs.\cite{xicc1,xicc2,xicc3,xicc4,xicc5,xicc6,xicc7,xicc8,xicc9}, where $Q^{(\prime)}$ stands for heavy $b$ or $c$ quark, respectively. For convenience, throughout the paper, we take $\Xi_{QQ'}$ as a short notation for the baryon $\Xi_{QQ'q}$, with $q$ equals to the light quark $u$, $d$, or $s$, respectively. Among the doubly heavy baryons $\Xi_{cc}$, $\Xi_{bc}$, and $\Xi_{bb}$, only $\Xi_{cc}$ has been observed by the SELEX fixed-target experiment~\cite{selex1,selex2}. However, the SELEX measurements on the $\Xi_{cc}$ properties, such as its decay width and production rate, are much larger than the theoretical predictions~\cite{arguSelex,selex1,selex2}, even by including the extrinsic and intrinsic charm production mechanisms~\cite{xicc7}. At present, its observations are also lack of supports from other experiments~\cite{lhcb,exp1,exp2}. Thus, in addition to the hadronic platforms, it is helpful to find other platforms which can generate large amounts of baryon events to study the baryon properties more precisely.

We shall study the doubly heavy baryon photoproduction in the future $e^+e^-$ International Linear Collider (ILC) within the framework of the non-relativistic QCD (NRQCD) theory~\cite{nrqcd}. Within this platform, the doubly heavy baryons can be produced through the channel via a single virtual photon or a $Z^0$ boson, $e^+ e^- \to \gamma^*/Z^0 \to \Xi_{QQ'} +\bar{Q'} +\bar{Q}$, or through the photonproduction channel via the double photon collision, $\gamma \gamma \to \Xi_{QQ'} +\bar{Q'} +\bar{Q}$. It is found that the production cross section for the single photon/$Z^0$ process shall be highly suppressed~\cite{xicc9}, which is about two orders lower than that of the photoproduction channel. So, in the present paper, we shall concentrate on the photoproduction channel.

One can treat the photoproduction of the baryon by two steps. The first step is for the incident photons coming from the electron and positron beams to produce the heavy $Q\bar{Q}$ and $Q'\bar{Q'}$ pairs. This step is pQCD calculable and can be treated by using the improved helicity amplitude approach~\cite{helicity}. The second step is that the heavy quarks $Q$ and $Q'$ evolving into a binding diquark $(QQ')$ with color- and spin- configuration $[n]$, e.g., $[n]=[^3S_1]_{\bar{\textbf{3}}}$, $[^1S_0]_{\textbf{6}}$ for $(cc)$ or $(bb)$ diquark and $(bc)_{\bar{\textbf{3}}}[^3S_1]$, $(bc)_{\textbf{6}}[^1S_0]$, $(bc)_{\bar{\textbf{3}}}[^1S_0]$, and $(bc)_{\textbf{6}}[^3S_1]$, respectively. Then, the diquark $(QQ')[n]$ shall be hadronized into the doubly heavy baryons $\Xi_{cc}$, $\Xi_{bc}$, and $\Xi_{bb}$ via fragmentation. Similar factorization procedures have also been suggested for dealing with the $\Lambda_c$ or $\Lambda_b$ baryon production~\cite{xicc10}.

According to NRQCD, the $\Xi_{QQ^{\prime}}$ baryon can be expanded over the Fock states,
\begin{eqnarray}
\vert\Xi_{QQ^{\prime}} \rangle &=& c_1(v) \vert (QQ^{\prime}) q\rangle +c_2(v) \vert (QQ^{\prime}) qg \rangle \nonumber\\
&& +c_3(v) \vert (QQ^{\prime}) q gg \rangle +\cdots, \nonumber
\end{eqnarray}
where $v$ is the relative velocity of the constituent heavy quarks in the baryon rest frame. Usually, it is stated that all the baryons are dominated by the first Fock state $|(QQ^{\prime})q\rangle$, then the emitted gluon from the heavy quark for $(QQ^{\prime})$ in $[^1S_0]_{\textbf{6}}$ state must change the spin of the heavy quark; Thus, the probability coefficient $c_1(v)$ shall dominant over other coefficients, or equivalently, $h_{\bf 6}$ shall be at least $v^2$-suppressed to $h_{\bf\bar{3}}$ and can be neglected. Here $h_{\bf\bar{3}}$ stands for the probability of transforming the color antitriplet diquark into the baryon and $h_{\bf 6}$ stands for the probability of transforming the color sextuplet diquark into the baryon.

A different power counting rule over $v$-expansion has also been suggested in the literature. Ref.\cite{xicc5} suggests that the second Fock state $|(QQ^{\prime}) qg \rangle$ can be of the same importance as $|(QQ^{\prime}) q \rangle$~\cite{xicc5}. Its main idea lies in that one of the heavy quarks can emit a gluon, which does not need to change the spin of the heavy quark, and this gluon can further split into a light $q\bar q$ pair; The light quarks can also emit gluons, and finally, the baryon components can be formed with a light quark $q$ plus one or more soft gluons. Since the light quark can emit gluons easily, we have $c_{1}(v)\sim c_{2}(v)\sim c_{3}(v)$. As a rough order estimation, we take the transition probabilities for those diquark states to form the corresponding baryon to be the same, i.e., $h_{\bf 6} \simeq h_{\bf\bar{3}}$. Following this approximation, we shall find that the color sextuplet diquark component can also provide sizable contributions to the baryon production. It is found that those matrix elements are overall parameters, and their uncertainties can be conveniently discussed when we know their values well. For convenience, we will adopt the assumption $h_{\bf 6} \simeq h_{\bf\bar{3}}$ to do our discussions throughout the paper.

The remaining parts of the paper are organized as follows. In Sec.II, we present the formulation for dealing with the photoproduction channel $\gamma \gamma \to \Xi_{QQ'} +\bar{Q'} +\bar{Q}$ at the leading-order level. In Sec.III, we give the numerical results. Sec.IV is reserved for a summary.

\section{Calculation technology}

At the leading order ${\cal O}(\alpha^2\alpha_s^2)$, within the NRQCD factorization approach, the differential cross section for the channel $\gamma \gamma \to \Xi_{QQ'} +\bar{Q'} +\bar{Q}$ can be formulated as
\begin{widetext}
\begin{equation}\label{corsec}
d\sigma = \int dx_1 dx_2 f_{\gamma}(x_1) f_{\gamma}(x_2) \int dz D^H_{QQ'}(z) \times \frac{1}{2 x_1 x_2 S} \overline{\sum}  |{\cal M}|^{2} d\Phi_3 \langle{\cal O}^H(n) \rangle\;,
\end{equation}
\end{widetext}
where $\langle{\cal O}^H(n) \rangle$ stands for the long-distance matrix element, which is proportional to the inclusive transition probability of the perturbative state, $(QQ')[n]$ pair into the heavy baryon $\Xi_{QQ'}$. ${\cal M}$ is the hard scattering amplitude, which is calculable since the intermediate gluon should be hard enough to generate a heavy $Q\bar{Q}$ or $Q'\bar{Q}'$ pair. $\overline{\sum}$ means we need to average over the spin states of the electron and positron and sum over the color and spin of all final particles. $d\Phi_3$ is the conventional three-body phase space. $f_{\gamma}(x)$ is the density function of the incident photon~\cite{ydenfun,lbs}.

\begin{figure*}[htb]
\includegraphics[width=1.0\textwidth]{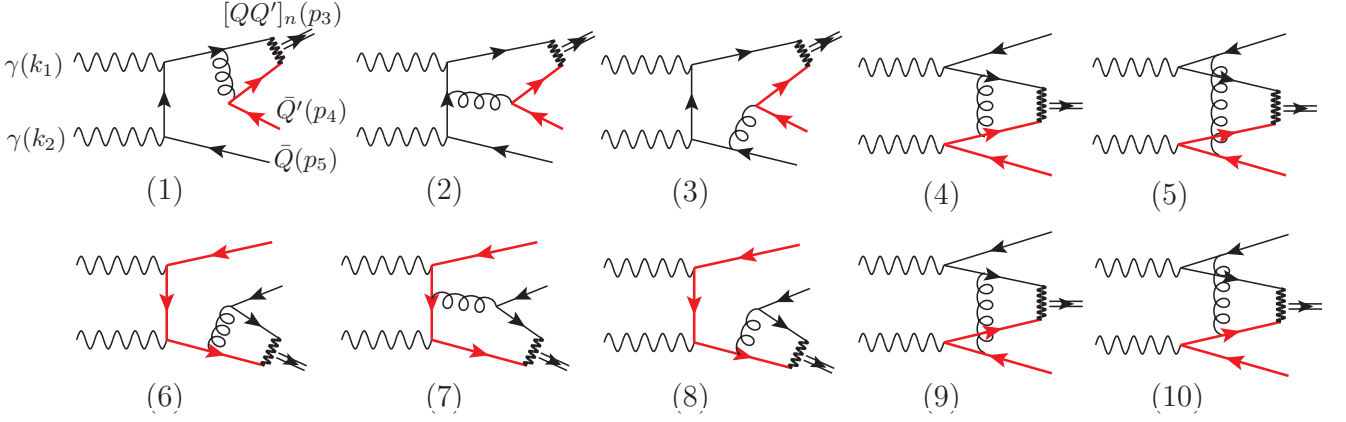}
\caption{Typical Feynman diagrams for the production channel $\gamma(k_{1})+\gamma(k_{2}) \to \Xi_{QQ'}(p_{3})+\bar{Q'}(p_4)+\bar{Q}(p_5)$ at the tree level. Other ten diagrams can be obtained by exchanging the positions of the initial photons attached to the quark lines.} \label{YYFeny}
\end{figure*}

There are totally 20 Feynman diagrams for the channel $\gamma(k_{1})+\gamma(k_{2}) \to \Xi_{QQ'}(p_{3}) +\bar{Q'}(p_4)+ \bar{Q}(p_5)$. We put 10 diagrams in Fig.~\ref{YYFeny}, the other 10 Feynman diagrams can be obtained by exchanging the positions of the initial photons attached to the quark lines. The fragmentation function $D_{QQ'}^{H}$ can be estimated within certain phenomenological models~\cite{frag1,frag2,frag3}. We adopt the following form suggested by Ref. \cite{frag1} for our calculation
\begin{eqnarray}
D_{QQ'}^{H}(z)=\frac{N_{QQ'}}{z[1-(1/z)-\epsilon_{QQ'}/(1-z)]^2},  \label{fragF}
\end{eqnarray}
where $\epsilon_{cc}=({m_q}/{M_{\Xi_{cc}}})^2 \epsilon_{b}$, $\epsilon_{bc} =({m_q}/{M_{\Xi_{bc}}})^2 \epsilon_{b}$, and $\epsilon_{bb}=({m_q}/{M_{\Xi_{bb}}})^2 \epsilon_{b}$. The light quark mass is chosen as $m_q=0.3$ GeV~\cite{Xicc98}. The parameter $\epsilon_{b}$ can be fixed by comparing with the data~\cite{exp}, which is $\sim0.004$~\cite{tetraquark}. The normalization factor $N_{QQ'}$ can be fixed by the normalization condition, $\int D_{QQ'}^{H}(z)dz=1$, which leads to $N_{cc}=0.0084$, $N_{bc}=0.00386$, and $N_{bb}=0.0025$ for $\Xi_{cc}$, $\Xi_{bc}$, and $\Xi_{bb}$, respectively.

The hard scattering amplitude ${\cal M}=\sum^{20}_{k=1}{\cal M}_{k}$ for the process can be written in a general form as
\begin{widetext}
\begin{eqnarray}\label{ampform}
{\cal M}_{k} &=& {\cal C}_{ijl} \times X_k \times \sum\limits_{s_1,s_3}\bar{u}_{s_1}(p_{32}) \Gamma_{1} s_{f}(k_{1},m_{Q'}) \cdots s_{f}(k_{\rho-1},m_{Q'}) \Gamma_{\rho} v_{s_2}(p_4) \nonumber \\
&& \times \bar{u}_{s_3}(p_{31}) \Gamma'_{1} s_{f}(k'_{1},m_{Q}) \cdots s_{f}(k'_{\kappa-1},m_{Q}) \Gamma'_{\kappa} v_{s_4}(p_5) ,
\end{eqnarray}
\end{widetext}
where $k=(1,\ldots,20)$, $\Gamma_1,\ldots,\Gamma_{\rho}$ and $\Gamma'_{1},\ldots,\Gamma'_{\kappa}$ are interaction vertexes, which contain the Dirac-$\gamma$ matrixes only. $s_{f}(k^{(_{'})}_{1},m_{Q^{(_{'})}})$ and the like are fermion propagators. $X_k$ is the scalar part of the propagators for the whole amplitude. The momentum of the constituent quarks are $p_{31}=\frac{m_Q}{M_{QQ'}}p_3$ and $p_{32}=\frac{m_{Q'}}{M_{QQ'}}p_3$. ${\cal C}_{ijl}$ is the color factor defined as
\begin{eqnarray}
{\cal C}_{ijl} &=& {\cal N}_c \times \sum\limits_{m,n}(T^a)_{im}(T^a)_{jn}\times G_{mnl},
\end{eqnarray}
where the subindices $m$ and $n$ are color indices of the constituent heavy quarks, and $l$ is the color of the diquark. $a=1,\ldots,8$ is the color index of the gluon propagator. ${\cal N}_c=1/\sqrt{2}$ is the normalization factor. The function $G_{mnl}$ equals the antisymmetric $\varepsilon_{mnl}$ (the symmetric $f_{mnl}$) for the color antitriplet $\bar{\textbf{3}}$ (the color sextuplet $\textbf{6}$) of $(QQ')$ diquark. The sum of the anti-symmetric and the symmetric functions satisfy the following equations
\begin{eqnarray}
\varepsilon_{mnl} \varepsilon_{m'n'l} = \delta_{mm'}\delta_{nn'}-\delta_{mn'}\delta_{nm'}
\end{eqnarray}
and
\begin{eqnarray}
f_{mnl} f_{m'n'l} = \delta_{mm'}\delta_{nn'}+\delta_{mn'}\delta_{nm'}.
\end{eqnarray}
With the help of the above relations, we obtain ${\cal C}_{ijl}^2=\frac{4}{3}$ for the color-antitriplet diquark state and ${\cal C}_{ijl}^2=\frac{2}{3}$ for the color-sextuplet diquark state, respectively.

All the amplitudes ${\cal M}_k$ with $k=(1,\ldots,20)$ contain massive quark lines, so it is too complicated and lengthy by using the conventional trace technique to deal with the amplitude square. To shorten the calculations and to make the results more compact, we adopt the improved helicity amplitude approach~\cite{helicity} to deal with the difficulty of calculating the expressions for the yields when the quark masses cannot be neglected. It is found that we can connect the doubly heavy baryon production with those of doubly heavy quarkonium production. We have made a detailed discussion on the heavy quarkonium production at the ILC under the improved helicity amplitude approach via the channel $\gamma\gamma \to |[Q\bar{Q'}](n)\rangle+Q'+\bar{Q}$ in Ref.~\cite{yyJpsi}. To compare with the quarkonium case, by applying the charge conjugation matrix $C=-i\gamma^2\gamma^0$ and the transverse of the matrix element to the amplitude ${\cal M}_k$, we can transform Eq.~(\ref{ampform}) as
\begin{widetext}
\begin{eqnarray}
{\cal M}_{k} &=& (-1)^{\rho+1} {\cal C}_{ij} \times X_k \times \sum\limits_{s_1,s_3} \bar{u}_{s_2}(p_4) \Gamma_{\rho} s_{f}(-k_{\rho-1},m_{Q'}) \cdots s_{f}(-k_{1},m_{Q'}) \Gamma_{1} v_{s_1}(p_{32}) \nonumber \\
&& \times \bar{u}_{s_3}(p_{31}) \Gamma'_{1} s_{f}(k'_{1},m_{Q}) \cdots s_{f}(k'_{\kappa-1},m_{Q}) \Gamma'_{\kappa} v_{s_4}(p_5) \nonumber\\
&=& (-1)^{\rho+1} {\cal C}_{ij} X_k \bar{u}_{s_2}(p_4) \Gamma_{\rho} s_{f}(-k_{\rho-1},m_{Q'}) \cdots s_{f}(-k_{1},m_{Q'}) \Pi(p_3) \Gamma'_{1} s_{f}(k'_{1},m_{Q}) \cdots s_{f}(k'_{\kappa-1},m_{Q}) \Gamma'_{\kappa} v_{s_4}(p_5),  \label{mt}
\end{eqnarray}
\end{widetext}
where $\rho$ stands for the number of the $\gamma$-matrixes appearing in the amplitude ${\cal M}_{k}$. The second line is the matrix element for the heavy quarkonium production, which indicates that the amplitudes for the diquark production are merely different from those of the heavy quarkonium case with an overall factor $(-1)^{\rho+1}$. Thus, inversely, we can conveniently derive the hard scattering amplitudes ${\cal M}_{k}$ from Ref.~\cite{yyJpsi} after proper transformation. To shorten the paper, we will not put the detailed calculation technology for the baryon production here, the interesting readers may turn to Ref.~\cite{yyJpsi} for details of the improved helicity amplitude approach. Here, to derive Eq.~(\ref{mt}), we have implicitly applied the relations: $CC^{-1}=1$ and
\begin{eqnarray}\label{operator}
&&v^T_{s}(p) C = -\bar{u}_{s}(p),\; C^{-1} \bar{u}^T_{s}(p)=v_{s}(p),\nonumber \\
 C^{-1} s^T_{f}&&(k_{1},m_{Q})C=s_{f}(-k_{1},m_{Q}),\; C^{-1}\Gamma^T_n C=-\Gamma_n.
\end{eqnarray}

\section{Numerical results and discussions}

As discussed in the Introduction, we adopt $h_{\bf 6} \simeq h_{\bf\bar{3}}$ to do our discussion. The nonperturbative matrix element with color antitriplet diquark, $h_{\bf\bar 3}$, can be related to the Schr\"{o}dinger wave functions at the origin $|\psi_{(Q\bar{Q'})}(0)|$ as~\cite{xicc5}:
\begin{eqnarray}
h_{\bf\bar{3}}=\langle{\cal O}^{H} (1 S) \rangle \simeq |\psi_{|(Q\bar{Q'})[1 S]\rangle}(0)|^2.
\end{eqnarray}
Since the spin-splitting effect is small, we do not distinguish the bound state parameters for the spin-singlet and the spin-triplet states; i.e., those parameters, such as the constituent quark masses, the bound state mass, and the wave function, are taken to be the same for the spin-singlet and spin-triplet states. We take the wavefunctions at the origin as~\cite{xicc3}: $|\Psi_{cc}(0)|^2 =0.039$ GeV$^3$, $|\Psi_{bc}(0)|^2=0.065$ GeV$^3$, and $|\Psi_{bb}(0)|^2 = 0.152$ GeV$^3$. The heavy quark masses are taken as: $m_c=1.5$ GeV and $m_b=4.9$ GeV. The doubly heavy baryon mass is taken as $M_{\Xi_{QQ'}}=m_{Q}+m_{Q'}$. The other parameters are taken as the same as those of Ref.~\cite{yyJpsi}, e.g., the renormalization scale is taken as the transverse mass, $\mu_r=M_t=\sqrt{M^2_{QQ'}+p_t^2}$. As a cross check of our calculation, we obtain same numerical results as those derived from the conventional squared amplitude approach.

\begin{table}[htb]
\begin{tabular}{|c|c|c|c|}
\hline
 ~~ & ~250 (GeV)~ & ~500 (GeV)~ & ~1 (TeV)~ \\
\hline
~~$(cc)_{\textbf{6}}[^1S_0]$~~ & 39.27 & 18.45 & 7.53 \\
\hline
$(cc)_{\bar{\textbf{3}}}[^3S_1]$ & 434.86 & 183.81 & 71.92 \\
\hline
$(bc)_{\bar{\textbf{3}}}[^3S_1]$ & 21.81 & 10.45 & 4.48 \\
\hline
$(bc)_{\textbf{6}}[^1S_0]$ & 4.81 & 2.27 & 0.96 \\
\hline
$(bc)_{\textbf{6}}[^3S_1]$ & 10.91 & 5.22 & 2.24 \\
\hline
$(bc)_{\bar{\textbf{3}}}[^1S_0]$ & 9.62 & 4.53 & 1.92 \\
\hline
$(bb)_{\textbf{6}}[^1S_0]$ & 0.04 & 0.02 & 0.01 \\
\hline
$(bb)_{\bar{\textbf{3}}}[^3S_1]$ & 0.53 & 0.28 & 0.13 \\
\hline
\end{tabular}
\caption{Total cross sections (in unit: fb) for the photoproduction of $\Xi_{cc}$, $\Xi_{bc}$, and $\Xi_{bb}$ under various color- and spin- configurations at the ILC. }\label{tcrs}
\end{table}

Total cross sections for the doubly heavy baryon photoproduction with three collision energies, i.e., $\sqrt{S}=250$ GeV, 500 GeV, and 1 TeV, are put in Table~\ref{tcrs}. Summing up the contributions from different color- and spin- configurations, we find that the total cross sections decrease with the increment of $\sqrt{S}$, i.e.,
\begin{eqnarray}
&& \sigma_{\Xi_{cc}}|_{250{\rm GeV}} : \sigma_{\Xi_{cc}}|_{500{\rm GeV}} : \sigma_{\Xi_{cc}}|_{1{\rm TeV}} \simeq 6:3:1 , \nonumber\\
&& \sigma_{\Xi_{bc}}|_{250{\rm GeV}} : \sigma_{\Xi_{bc}}|_{500{\rm GeV}} : \sigma_{\Xi_{bc}}|_{1{\rm TeV}} \simeq 5:2:1 , \nonumber\\
&& \sigma_{\Xi_{bb}}|_{250{\rm GeV}} : \sigma_{\Xi_{bb}}|_{500{\rm GeV}} : \sigma_{\Xi_{bb}}|_{1{\rm TeV}} \simeq 4:2:1 . \nonumber
\end{eqnarray}
It is noted that the relative importance among different color- and spin- configurations for the total cross sections and the differential distributions are similar under different collision energies. In the following, we take $\sqrt{S}=500\;{\rm GeV}$ as the $e^+e-$ collision energy.

Under the condition of $\sqrt{S}=500\;{\rm GeV}$, we obtain
\begin{eqnarray}
&&\sigma_{(cc)_{\bar{\textbf{3}}}[^3S_1]} : \sigma_{(cc)_{\textbf{6}}[^1S_0]} \simeq 10:1 , \nonumber\\
&&\sigma_{(bc)_{\bar{\textbf{3}}}[^3S_1]} : \sigma_{(bc)_{\textbf{6}}[^1S_0]} : \sigma_{(bc)_{\textbf{6}}[^3S_1]} : \sigma_{(bc)_{\bar{\textbf{3}}}[^3S_1]}  \simeq 5:1:2:2 , \nonumber \\
&&\sigma_{(bb)_{\bar{\textbf{3}}}[^3S_1]} : \sigma_{(bb)_{\textbf{6}}[^1S_0]} \simeq 14:1 . \nonumber
\end{eqnarray}
It indicates that the $[^3S_1]_{\bar{\textbf{3}}}$ diquark state provides the dominant contribution, while other configurations may also provide significant contributions. By summing up all the possible diquark configurations, we obtain $\sigma_{\Xi_{cc}}=202.26$ fb, $\sigma_{\Xi_{bc}}=22.47$ fb, and $\sigma_{\Xi_{bb}}=0.3$ fb. If the integrated luminosity is as high as $10^{4}$ fb$^{-1}$, we shall have about $2.0\times 10^{6}$ $\Xi_{cc}$, $2.2\times 10^{5}$ $\Xi_{bc}$, and $3.0\times 10^{3}$ $\Xi_{bb}$ events to be generated through the direct photon collision at the ILC in an operation year. The $\Xi_{cc}$ production rate is larger than those of $\Xi_{bc}$ and $\Xi_{bb}$, i.e., $\sigma_{\Xi_{cc}}:\sigma_{\Xi_{bc}}:\sigma_{\Xi_{bb}}=647:75:1$. Thus, in the following, we shall focus on the photoproduction of $\Xi_{cc}$ and $\Xi_{bc}$.

\begin{figure}[htb]
\includegraphics[width=0.48\textwidth]{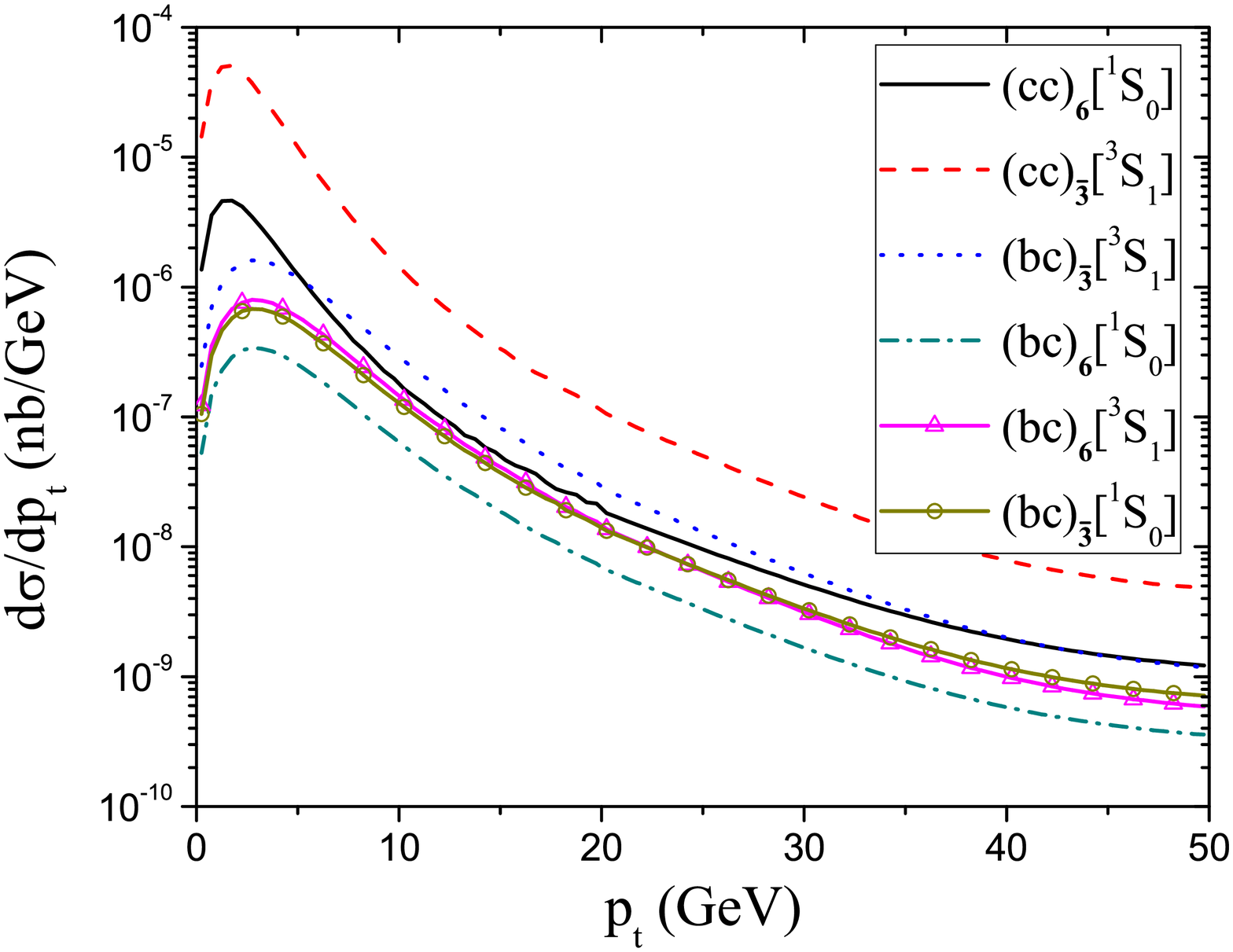}
\caption{The baryon $p_t$ distributions for the process $\gamma\gamma \to \Xi_{QQ'}+ \bar{Q'}+\bar{Q}$ at the ILC with $\sqrt{S}=500$ GeV, where the production via different $(QQ')[n]$-diquark configurations are presented. } \label{pt}
\end{figure}

\begin{figure}[htb]
\includegraphics[width=0.48\textwidth]{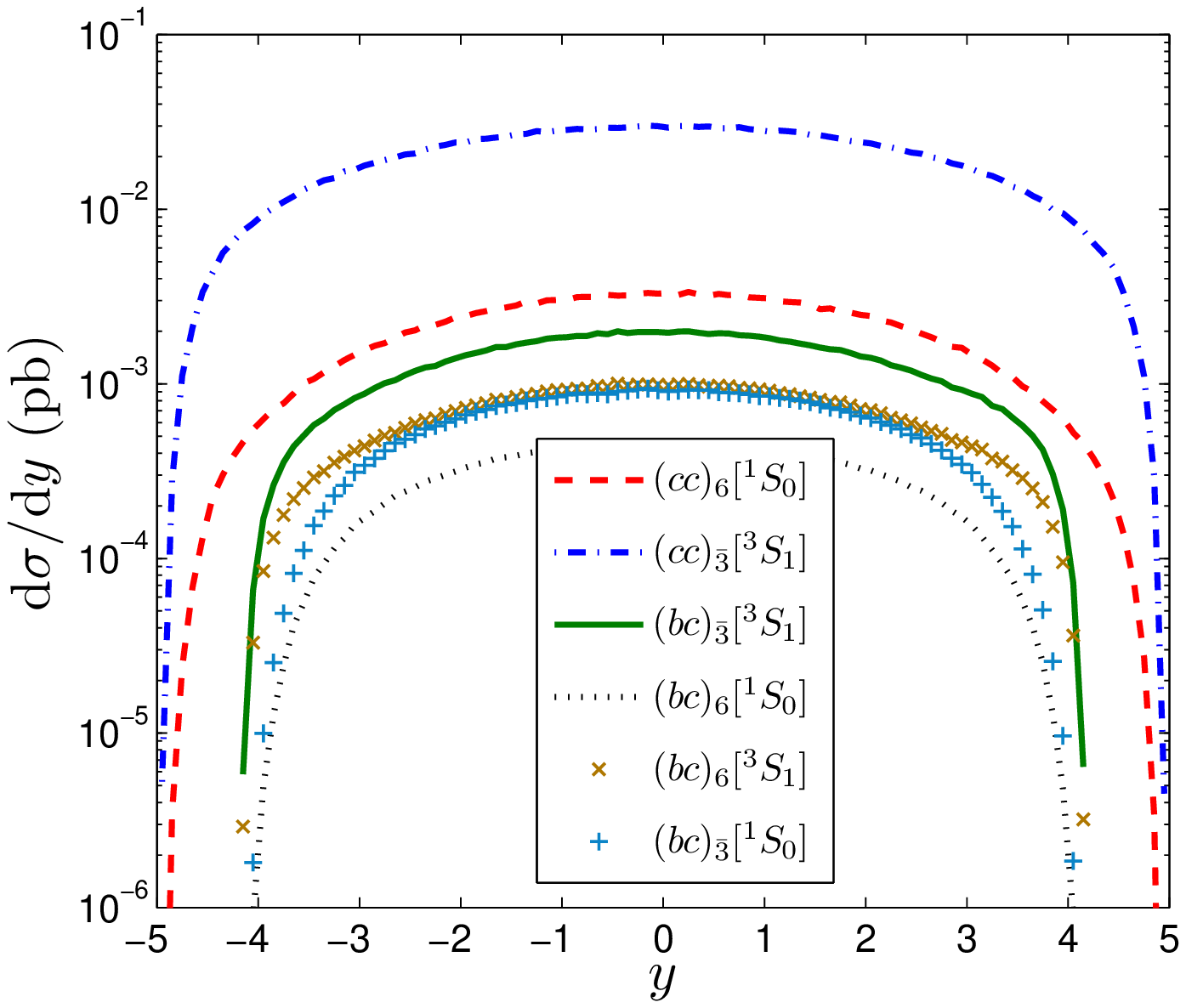}
\caption{The baryon rapidity distributions for the process $\gamma\gamma \to \Xi_{QQ'}+ \bar{Q'}+\bar{Q}$ at the ILC with $\sqrt{S}=500$ GeV, where the production via different $(QQ')[n]$-diquark configurations are presented. } \label{rap}
\end{figure}

\begin{figure}[htb]
\includegraphics[width=0.48\textwidth]{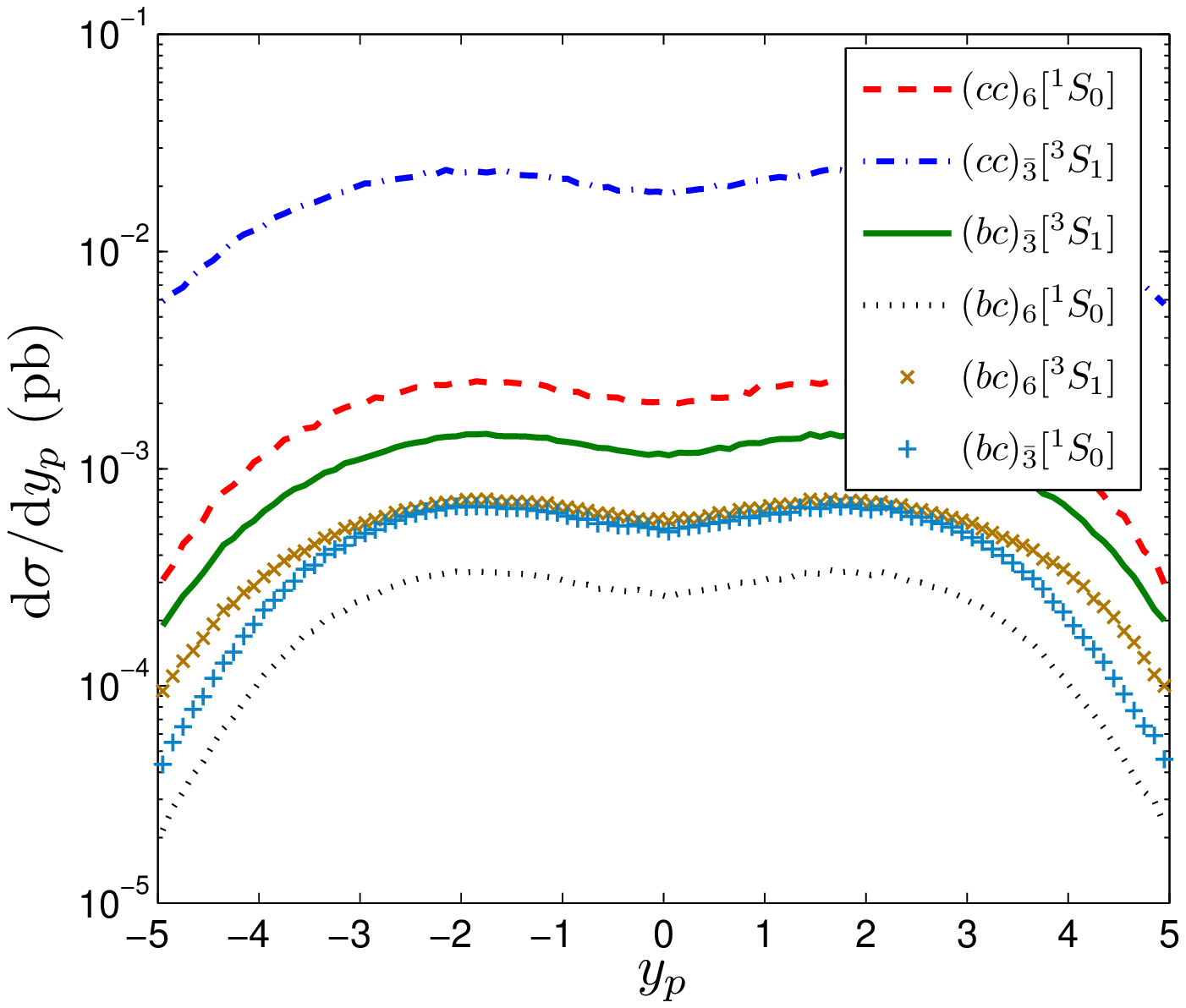}
\caption{The baryon pseudorapidity distributions for the process $\gamma\gamma \to \Xi_{QQ'}+ \bar{Q'}+\bar{Q}$ at the ILC with $\sqrt{S}=500$ GeV, where the production via different $(QQ')[n]$-diquark configurations are presented. } \label{psrap}
\end{figure}

\begin{figure}[htb]
\includegraphics[width=0.48\textwidth]{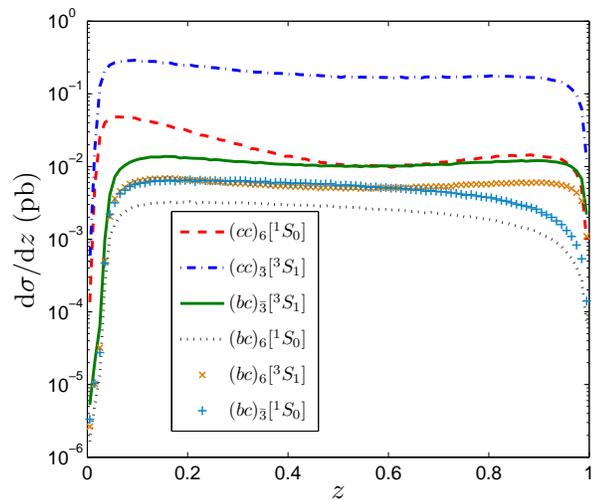}
\caption{The differential distributions $d\sigma/dz$ for the process $\gamma\gamma \to \Xi_{QQ'}+ \bar{Q'}+\bar{Q}$ at the ILC with $\sqrt{S}=500$ GeV, where the production via different $(QQ')[n]$-diquark configurations are presented. }  \label{yyz}
\end{figure}

\begin{table}[htb]
\begin{tabular}{|c|c|c|c|}
\hline
~~$p_{t\textrm{-cut}}$~~ & ~1 GeV~ & ~2 GeV~ & ~3 GeV~  \\
\hline
~~$(cc)_{\textbf{6}}[^1S_0]$~~ & 16.11 & 11.47 & 7.81\\
\hline
$(cc)_{\bar{\textbf{3}}}[^3S_1]$& 159.24 & 112.44 & 71.25 \\
\hline
$(bc)_{\bar{\textbf{3}}}[^3S_1]$& 10.03 & 8.85 & 7.35 \\
\hline
$(bc)_{\textbf{6}}[^1S_0]$& 2.17 & 1.92 & 1.60   \\
\hline
$(bc)_{\textbf{6}}[^3S_1]$& 5.01 & 4.42 &  3.67 \\
\hline
$(bc)_{\bar{\textbf{3}}}[^1S_0]$& 4.34 & 3.84 & 3.20 \\
\hline
\end{tabular}
\caption{Total cross sections (in units fb) for the photoproduction of $\Xi_{cc}$ and $\Xi_{bc}$ with $\sqrt{S}=500$ GeV under various color- and spin- configurations and various $p_t$ cuts.}\label{ptcuts}
\end{table}

\begin{table}[htb]
\begin{tabular}{|c|c|c|c|}
\hline
~~$y_{\textrm{cut}}$~~ & ~~~1~~~ & ~~~2~~~ & ~~~3~~~ \\
\hline
~~$(cc)_{\textbf{6}}[^1S_0]$~~& 5.85 & 11.53 & 15.76 \\
\hline
$(cc)_{\bar{\textbf{3}}}[^3S_1]$& 53.9 & 105.2 & 148.1 \\
\hline
$(bc)_{\bar{\textbf{3}}}[^3S_1]$& 3.76 & 7.02 & 9.35   \\
\hline
$(bc)_{\textbf{6}}[^1S_0]$& 0.85 & 1.61 & 2.12 \\
\hline
$(bc)_{\textbf{6}}[^3S_1]$& 1.88 & 3.51 & 4.67 \\
\hline
$(bc)_{\bar{\textbf{3}}}[^1S_0]$& 1.70 & 3.22 & 4.24 \\
\hline
\end{tabular}
\caption{Total cross sections (in units fb) for the photoproduction of $\Xi_{cc}$ and $\Xi_{bc}$ with $\sqrt{S}=500$ GeV under various color- and spin- configurations and various rapidity cuts.}\label{ycuts}
\end{table}

Fig.~\ref{pt} shows the baryon transverse momentum $(p_t)$ distributions for the photoproduction of $\Xi_{cc}$ and $\Xi_{bc}$. Similar to the above conclusion, the $[^3S_1]_{\bar{\textbf{3}}}$ configuration for both $\Xi_{cc}$ and $\Xi_{bc}$ production provides dominant contributions over the other configurations in the whole $p_t$ region. We present the rapidity ($y$) and pseudorapidity ($y_p$) distributions in Figs.~\ref{rap} and \ref{psrap}. There is a plateau within $|y|<4$ or $|y_p|<4$. We present the differential cross sections $d\sigma/dz$ in Fig.~\ref{yyz}, where $z =\frac{2}{\hat{s}}(k_1+k_2)\cdot p_3$ with $\hat{s}=x_1 x_2 S$ being the invariant mass of the initial photons of the subprocess. In the subprocess center-of-mass frame, $z$ is simply twice the fraction of the total energy carried by the baryon and is experimentally observable. To be useful references, we present the total cross sections under various $p_t$ or $y$ cuts in Tables \ref{ptcuts} and \ref{ycuts}.

\begin{table}[htb]
\begin{tabular}{|c|c|c|}
\hline
 & ~~~$\sigma_{\textrm{d}}$~~~ & ~~~$\sigma_{\textrm{f}}$~~~ \\
\hline
~$(cc)_{\textbf{6}}[^1S_0]$~ & 18.41 & 18.45 \\
\hline
$(cc)_{\bar{\textbf{3}}}[^3S_1]$ & 184.62 & 183.81 \\
\hline
$(bc)_{\bar{\textbf{3}}}[^3S_1]$ & 10.49 & 10.45 \\
\hline
$(bc)_{\textbf{6}}[^1S_0]$ & 2.28 & 2.27 \\
\hline
$(bc)_{\textbf{6}}[^3S_1]$ & 5.24 & 5.22 \\
\hline
$(bc)_{\bar{\textbf{3}}}[^1S_0]$ & 4.56 & 4.53 \\
\hline
\end{tabular}
\caption{Comparison of the total cross sections (in units fb) for the baryon photoproduction at the ILC with $\sqrt{S}=500$ GeV. The subscript ``d" stands for the ``direct evolution", the subscript ``f" stands for the ``evolution via fragmentation". } \label{compcs}
\end{table}

\begin{figure}[htb]
\includegraphics[width=0.48\textwidth]{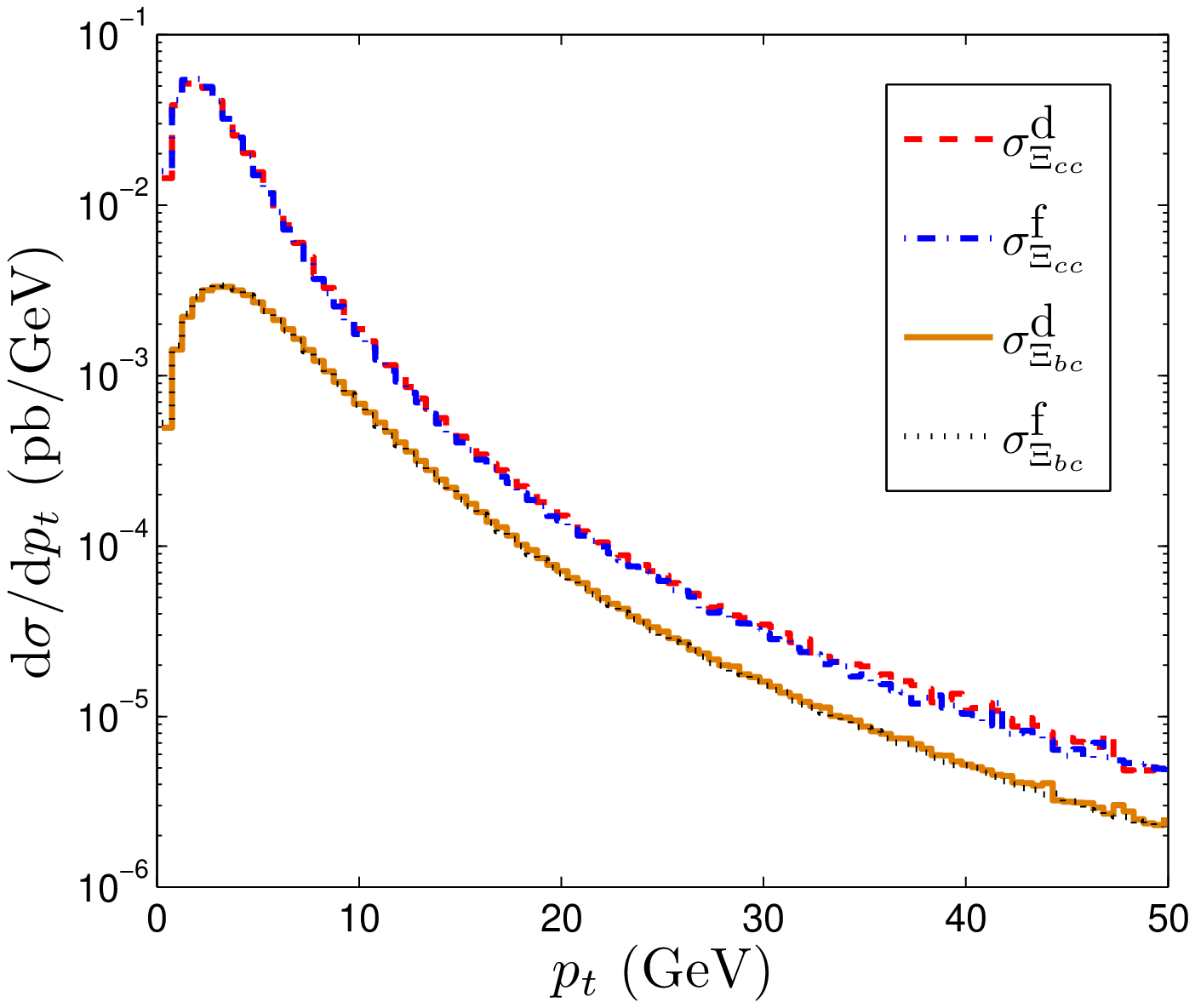}
\caption{Comparison of the $p_t$ distributions for the photoproduction of the $\Xi_{cc}$ and $\Xi_{bc}$ baryons at the ILC with $\sqrt{S}=500$ GeV. The superscript ``d" stands for the ``direct evolution", the superscript ``f" stands for the ``evolution via fragmentation". }\label{comppt}
\end{figure}

\begin{figure}[htb]
\includegraphics[width=0.48\textwidth]{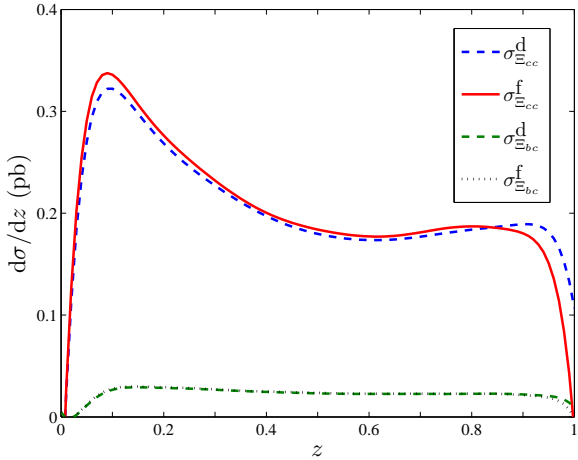}
\caption{Comparison of the $z$ distributions for the photoproduction of the $\Xi_{cc}$ and $\Xi_{bc}$ baryons at the ILC with $\sqrt{S}=500$ GeV. The superscript ``d" stands for the ``direct evolution", the superscript ``f" stands for the ``evolution via fragmentation". }\label{compz}
\end{figure}

In the literature, people usually takes a simple assumption by treating the evolution from the diquark to doubly heavy baryon with $100\%$ probability and with equal importance for all phase-space point; we call it the ``direct evolution". In the present paper, we have adopted the fragmentation approach with the help of the fragmentation function (\ref{fragF}) to deal with such evolution; we call it the ``evolution via fragmentation". In Table \ref{compcs}, we present a comparison of the total cross sections for the baryon photoproduction at the ILC  under those two treatments. The subscript ``d" stands for the ``direct evolution", the subscript ``f" stands for the ``evolution via fragmentation". Table \ref{compcs} shows the discrepancies of total cross sections for those two treatments are quite small, i.e., less than $\sim1\%$. On the other hand, the differences for the $p_t$ distributions are also very small in the whole $p_t$ region. For example, we put a comparison of the $p_t$ distributions under those two treatments in Fig.~\ref{comppt}. We also present a comparison of the $z$ distributions under those two treatments in Fig.~\ref{compz}. In those two figures, we have summed up the contributions from the mentioned color- and spin- diquark-configurations for convenience. As for the $\Xi_{cc}$ production, it is found that the differential cross sections for the ``evolution via fragmentation" are slightly larger in small $z$ region, while slightly smaller in large $z$ region. This result is consistent with the small differences for the $p_t$ distributions under those two treatments. Thus the conventional treatment is viable and provides a good approximation to deal with the heavy baryon production. As a sound estimation, we take the fragmentation approach to do the discussion.

\begin{table}[htb]
\begin{tabular}{|c|c|c|c|}
\hline
~~$m_c$ (GeV)~~ & ~~~1.4~~~ & ~~~1.5~~~ & ~~~1.6~~~ \\
\hline
~$(cc)_{\textbf{6}}[^1S_0]$~ & 24.68 & 18.45 & 13.88\\
\hline
$(cc)_{\bar{\textbf{3}}}[^3S_1]$& 245.38 & 183.81 & 141.30  \\
\hline
$(bc)_{\bar{\textbf{3}}}[^3S_1]$& 12.23 & 10.45 & 9.08 \\
\hline
$(bc)_{\textbf{6}}[^1S_0]$& 2.62 & 2.27 & 1.98 \\
\hline
$(bc)_{\textbf{6}}[^3S_1]$& 6.12 & 5.22 & 4.54 \\
\hline
$(bc)_{\bar{\textbf{3}}}[^1S_0]$& 5.24 & 4.45 & 3.96 \\
\hline
\end{tabular}
\caption{Uncertainties for the total cross sections (in units fb) by taking $m_c=1.5\pm0.1$ GeV. $m_b=4.9$ GeV and $\mu_r=M_t$. }  \label{unmc}
\end{table}

\begin{table}[htb]
\begin{tabular}{|c|c|c|c|}
\hline
~~$m_b$ (GeV)~~ & ~~~4.7~~~ & ~~~4.9~~~ & ~~~5.1~~~ \\
\hline
$(bc)_{\bar{\textbf{3}}}[^3S_1]$& 11.37 & 10.45 & 9.67   \\
\hline
$(bc)_{\textbf{6}}[^1S_0]$& 2.46 & 2.27 & 2.07 \\
\hline
$(bc)_{\textbf{6}}[^3S_1]$& 5.68 & 5.22 & 4.83 \\
\hline
$(bc)_{\bar{\textbf{3}}}[^1S_0]$& 4.92 & 4.53 & 4.14 \\
\hline
\end{tabular}
\caption{Uncertainties for the total cross sections (in units fb) by taking $m_b=4.9\pm0.2$ GeV. $m_c=1.5$ GeV and $\mu_r=M_t$. } \label{unmb}
\end{table}

As a final remark, we make a discussion on the theoretical uncertainties from the heavy quark masses. For the purpose, we set $m_c=1.50\pm0.10$ GeV and $m_b=4.9\pm0.20$ GeV. As shown in Table \ref{unmc}, the uncertainties for $m_c=1.50\pm0.10$ GeV are
\begin{eqnarray}\label{cmass}
\sigma_{(cc)_{\textbf{6}}[^1S_0]} &=& 18.45^{+6.23}_{-4.57}\;{\rm fb},\nonumber\\
\sigma_{(cc)_{\bar{\textbf{3}}}[^3S_1]} &=& 183.81^{+61.57}_{-42.51}\;{\rm fb},\nonumber\\
\sigma_{(bc)_{\bar{\textbf{3}}}[^3S_1]} &=& 10.45^{+1.78}_{-1.37}\;{\rm fb},\nonumber\\
\sigma_{(bc)_{\textbf{6}}[^1S_0]} &=& 2.27^{+0.35}_{-0.29}\;{\rm fb},\nonumber\\
\sigma_{(bc)_{\textbf{6}}[^3S_1]} &=& 5.22^{+0.89}_{-0.68}\;{\rm fb},\nonumber\\
\sigma_{(bc)_{\bar{\textbf{3}}}[^1S_0]} &=& 4.53^{+0.70}_{-0.58}\;{\rm fb}.
\end{eqnarray}
Similarly, as shown in Table \ref{unmb}, the uncertainties caused by the $b$-quark mass $m_b=4.9\pm0.20$ GeV are
\begin{eqnarray}\label{bmass}
\sigma_{(bc)_{\bar{\textbf{3}}}[^3S_1]} &=& 10.45^{+0.92}_{-0.78}\;{\rm fb},\nonumber\\
\sigma_{(bc)_{\textbf{6}}[^1S_0]} &=& 2.27^{+0.19}_{-0.20}\;{\rm fb},\nonumber\\
\sigma_{(bc)_{\textbf{6}}[^3S_1]} &=& 5.22^{+0.46}_{-0.39}\;{\rm fb},\nonumber\\
\sigma_{(bc)_{\bar{\textbf{3}}}[^1S_0]} &=& 4.53^{+0.38}_{-0.40}\;{\rm fb}.
\end{eqnarray}

\begin{table}
\begin{tabular}{|c|c|c|c|}
\hline
~~$\mu_r$~~ & ~~~$\sqrt{\hat{s}}$~~~ & ~~~$\sqrt{\hat{s}}/2$~~~ & ~~~$M_t$~~~ \\
\hline
~$(cc)_{\textbf{6}}[^1S_0]$~ & 11.29 & 12.87 & 18.45 \\
\hline
$(cc)_{\bar{\textbf{3}}}[^3S_1]$& 117.67 & 135.53 & 183.81 \\
\hline
$(bc)_{\bar{\textbf{3}}}[^3S_1]$& 7.67 & 8.64 & 10.45 \\
\hline
$(bc)_{\textbf{6}}[^1S_0]$& 1.68 & 1.88 & 2.27 \\
\hline
$(bc)_{\textbf{6}}[^3S_1]$& 3.83 & 4.32 & 5.22 \\
\hline
$(bc)_{\bar{\textbf{3}}}[^1S_0]$& 3.36 & 3.76 & 4.53 \\
\hline
\end{tabular}
\caption{Total cross sections (in units fb) for the heavy quarkonium photoproduction under the improved conventional renormalization scale setting for three scale choices $\mu_r=\sqrt{\hat{s}}$, $\sqrt{\hat{s}}/2$, and $M_t$. $\sqrt{S}=500$ GeV. } \label{imprQ}
\end{table}

We take three scales $\mu_{r}=M_t$, $\sqrt{\hat{s}}/2$, and $\sqrt{\hat{s}}$ for estimating the scale uncertainties. Numerical results are shown in Table \ref{imprQ}, which indicates that the scale uncertainties are $\sim36\%$ for $\Xi_{cc}$ and $\sim26\%$ for $\Xi_{bc}$. Here we have adopted the improved way as suggested by Ref.~\cite{pmc} to analyze the scale uncertainty.

\section{Summary}

We have investigated the photoproduction of the doubly heavy baryons at the ILC within NRQCD. The improved helicity amplitude approach has been adopted to improve the calculation efficiency. By taking the assumption, $h_{\bf 6} \simeq h_{\bf\bar{3}}$, we observe that the channel via the intermediate $[^3S_1]_{\bar{\textbf{3}}}$ diquark state provides the dominant contribution, while other configurations may also provide significant contributions. Total and differential cross sections, together with their theoretical uncertainties, have been presented. By taking the errors from the heavy quark masses into consideration, we shall have $\left(2.0^{+0.68}_{-0.47}\right)\times 10^{6} \; \Xi_{cc}$ and $\left(2.2^{+0.37}_{-0.29}\right)\times 10^{5} \; \Xi_{bc}$ events to be produced in one operation year at the ILC with $\sqrt{S}=500$ GeV and ${\cal L}\simeq 10^{36}$cm$^{-2}$s$^{-1}$. Thus, the ILC would provide another good platform for studying $\Xi_{QQ'}$-baryon properties.

As a final remark, we discuss the possibility of distinguishing the baryon with different light constituent quark. As suggested by PYTHIA~\cite{pythia}, the relative probability for various doubly heavy baryons is $\sigma_{\Xi_{QQ'u}}: \sigma_{\Xi_{QQ'd}}: \sigma_{\Xi_{QQ's}}=10:10:3$. Then, for the produced $\Xi_{cc}$ events, one expects $43\%$ to be $\Xi_{cc}^{++}$, $43\%$ to be $\Xi_{cc}^+$, and $14\%$ to be $\Omega_{cc}^+$. The same situation occurs for the production of $\Xi_{bb}^0$, $\Xi_{bb}^-$, $\Omega_{bb}^-$, and $\Xi_{bc}^+$, $\Xi_{bc}^0$, $\Omega_{bc}^0$.

\hspace{2cm}

{\bf Acknowledgement:} This work was supported in part by the Fundamental Research Funds for the Central Universities under Grant No.CQDXWL-2012-Z002 and by the Natural Science Foundation of China under Grant No.11275280.

\end{document}